\documentstyle[
preprint,aps,graphicx
,epsfig
]{revtex}
 \tightenlines

\def\ket#1{\mathinner{|{#1}\rangle}}

{\catcode`\|=\active
  \gdef\Braket#1{\left<\mathcode`\|"8000\let|\bravert {#1}\right>}}
\def\bravert{\egroup\,\vrule\,\bgroup}

\begin{document}

\title{Non-deterministic Gates for Photonic
Single Rail Quantum Logic}
\author{A.~P.~Lund and T.~C.~Ralph}
\address{Centre for Quantum Computer Technology, Department of Physics
\\ University of Queensland, QLD 4072, Australia
\\ Fax: +61 7 3365 1242  Telephone: +61 7 3365 3412 \\
email: lund@physics.uq.edu.au\\}

\maketitle

\begin{abstract}

We discuss techniques for producing, manipulating and measureing 
qubits encoded optically as vacuum and single photon states. We show 
that a universal set of non-deterministic gates can be constructed 
using linear optics and photon counting. 
We investigate the efficacy of a test gate given realistic 
detector efficiencies.

\end{abstract}


\section{Introduction}

The standard method for encoding qubits in optics is to use the polarisation
degrees of freedom of single photons. This is sometimes referred to as dual
rail
logic because it uses the occupation, or not, of two orthogonal polarisation
modes as the qubit \cite{mil88}. This type of encoding is easy to
manipulate at
the single qubit level and recently schemes for two-qubit operations have been
introduced \cite{kni00,ral01,kni01}.
However, there is considerable fundamental interest in alternate encoding
strategies.

One such strategy is single rail logic \cite{lee01}. Here the
qubit is encoded in the occupation, or not, of a single optical mode. That is
the vacuum state, $\ket{0}$ represents the logical zero, whilst the single
photon
state, $\ket{1}$, represents the logical one. Recently experimental
progress in
creating superpositions of such states has been reported
\cite{res02,lvo02} and there has been a demonstration of entanglement
swapping based on this logic \cite{lom02}.

In this paper we show how it is possible to construct a universal set
of non-deterministic quantum gates for this
encoding using only linear optics and detection.
We investigate some simple experimental arrangements designed to test the
performance of the gates and conclude that demonstrations using state of the
art technology are possible. It would presumably be possible to scale up these
gates into near deterministic gates using similar techniques to those
proposed in Ref.\cite{kni00}, though we do not pursue this possibility
here. Alternatively the experiments proposed here may be
viewed as stepping stones to multi-photon single rail schemes, for
which scalable achitectures have been described \cite{ral02}.

A universal set of gates is formed by the Control sign shift (CS)
gate, the Hadamard gate and the phase rotation gate.
Non-deterministic CS gates for single rail logic
actually form the
heart of identical gates for dual rail logic \cite{kni00,ral01}.
We will borrow the most efficient and dedicated version of these,
recently described by Knill \cite{kni01}, for this discussion. Phase
rotations are easily implemented via phase delays but the Hadamard
gate presents a bigger challenge. Never-the-less we show that a
non-deterministic Hadamard gate can be implemented using only linear
optics and detection.

The paper is arranged in the following way. In the following section
we will review the construction of a CS gate. In section III we will
describe the construction of the Hadamard gate. In section IV we will
investigate the operation of our gates under non-ideal conditions and
in section V we will conclude.

\section{CS Gate}

We now review the operation of the linear optical network proposed by  Knill
\cite{kni01} and shown in Fig.1. The network is designed to be a
non-deterministic CS gate with a probability of success of $2/27$.

\noindent In this gate the reflectivities of the beam-splitters are
$\eta_1 = 1/3$ and $\eta_2 = \frac{1}{6}(3+\sqrt{6})$.  Also, these 
beam-splitters
have a sign change on transmission for a beam incident on the black side.
Therefore the operator evolution through individual beam-splitters looks like

\[
a_{out} = \sqrt{\eta}\ a + \sqrt{1-\eta}\ b
\]
\[
b_{out} = -\sqrt{1-\eta}\ a + \sqrt{\eta}\ b
\]

\noindent
and the operator evolution through the entire gate with the above
reflectivities is given by

\begin{eqnarray}
a_{out} & = & \frac{1}{3}(-a + \sqrt{2}b + \sqrt{2}c - 2d) \label{opstart} \\
b_{out} & = & \frac{1}{3}(-\sqrt{2}a - b +2c + \sqrt{2}d) \\
c_{out} & = &
\frac{1}{3\sqrt{2}}\left(\sqrt{3+\sqrt{6}}\left(\sqrt{2}a+c\right) +
\sqrt{3 - \sqrt{6}}\left(\sqrt{2}b + d\right)\right) \\
d_{out} & = &
\frac{1}{3\sqrt{2}}\left(-\sqrt{3-\sqrt{6}}\left(\sqrt{2}a+c\right)
 + \sqrt{3 + \sqrt{6}}\left(\sqrt{2}b + d\right)\right) \label{op} 
\end{eqnarray}

\noindent
where the qubits enter in modes $a$ and $b$, whilst modes $c$ and $d$
are prepared in single photon Fock states.
The input state to the gate can be written in a general way as
\begin{equation}
|\phi \rangle_{in} = c^{\dagger} d^{\dagger}(\alpha+\beta
a^{\dagger}+\gamma b^{\dagger}+\delta a^{\dagger}b^{\dagger}) |0000
\rangle
\label{csin}
\end{equation}
The output state of the gate can be calculated by inverting the
operator equations, Eq.\ref{opstart}-\ref{op}, thus obtaining the input
operators in
terms of the output operators, and substituting these for the
operators appearing in Eq.\ref{csin}. The form of the output state
is quite long and hence will not be given here.  However if we
condition our state by the simultaneous detection of a single photon
in each of the lower two modes $c$ and $d$, then the output state
(unnormalized) is
\begin{equation}
|\phi \rangle_{out} = \sqrt{{{2}\over{27}}}(\alpha+\beta
a_{o}^{\dagger}+\gamma b_{o}^{\dagger}-\delta a_{o}^{\dagger}b_{o}^{\dagger})
|0000 \rangle
\label{csout}
\end{equation}

\noindent This is control sign logic with a 180$^{\circ}$ sign flip.
The probability of success is $\frac{2}{27}$ as predicted
by Knill.

%
\begin{figure}[ht]
   \begin{center}
   \includegraphics[width=0.9\linewidth]{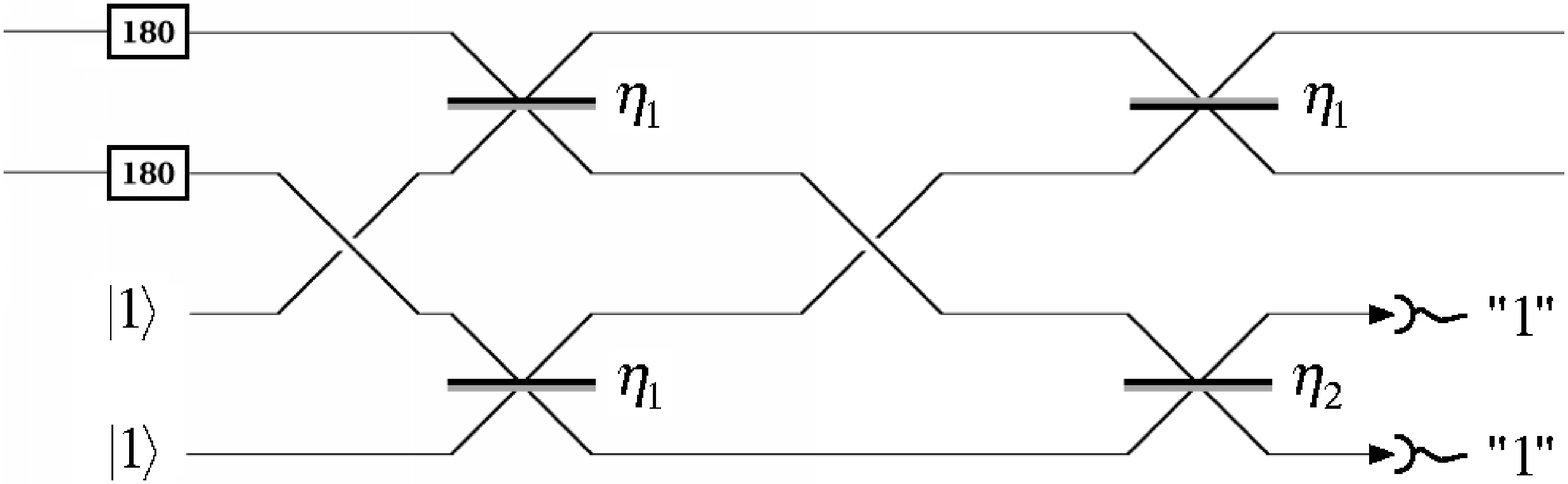}
   \end{center}
   \caption{Schematic of the CS gate proposed by Knill \protect\cite{kni01}.  
	The beam-splitters have a sign change upon 	
	transmission for light incident upon the black side.  This sign 
	convention is used for equations~\ref{opstart}-\ref{op}.
	The reflectivities of the beam-splitters are 
	$\eta_1 = 1/3$ and $\eta_2 = \frac{1}{6}(3+\sqrt{6})$.}	
   \label{CS}
\end{figure}
%

\section{Hadamard Gate}

{\it Superposition State Production}
In order to create the single rail Hadamard gate, a special superposition
is used and this
resource must be generated. The state that is desired is the equal
superposition state $\ket{0}+\ket{1}$.
This state can be produced conditionally using only linear optics
with coherent state
and single photon state inputs \cite{bar98}. We consider a simpler,
single element approach similar to that employed in the experiment of
Ref.\cite{lvo02}. This
set-up is shown Fig.2.
%
\begin{figure}[ht]
   \begin{center}
   \includegraphics[width=0.65\linewidth]{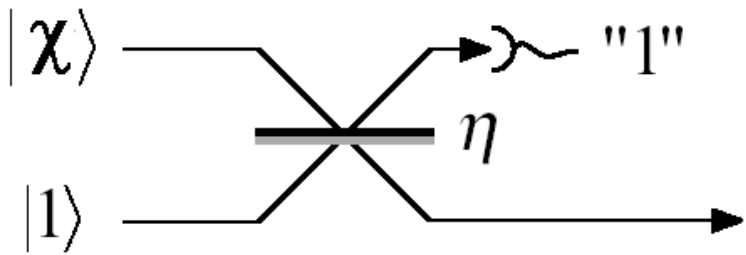}
   \end{center}
   \caption{Schematic of superposition production aparatus.  $\ket{\chi}$ here 
	is a coherent state with amplitude $\chi$.  The output (lower) mode of 
	this device will contain a superposition $\ket{0} + \ket{1}$ with 
	small higher order terms after the correct detection event has occured.}
   \label{bs}
\end{figure}

\noindent The parameter for the coherent state throughout will be given by
the number $\chi$.  This parameter and the reflectivity of the beam-splitter
must be chosen such as to give the $\ket{0}+\ket{1}$ state with the higher
order terms giving very little contribution to the state.  We
must assume that $\chi$ is close to zero to satisfy this situation.
The
analysis
of this system was performed using the expansion of the coherent state to
the order of three photons.  The coherent state $\ket{\chi}$ was written as
(unnormalised)

\[
\ket{\chi} \approx \ket{0} + \chi \ket{1} + \sqrt{\frac{1}{2}}\chi^2 \ket{2} +
\sqrt{\frac{1}{6}}\chi^3 \ket{3}\ \textrm{.}
\]

\noindent After the detection of one photon at the indicated output, the
following reflectivity ($\eta$) is required for the coefficients of $\ket{0}$
and $\ket{1}$ to be equal at the other output.

\begin{equation}
\eta = \frac{-1 + 4 \chi^2 \pm \sqrt{1+8\chi^2}}{8 \chi^2}
\label{rf}
\end{equation}

\noindent The positive solution is used from here on.  Using this
relationship, the value of $\chi$ required to make the coefficient of the
second order $\ket{2}$ term 100 times smaller (probability 10000 times
smaller) is

\[
\chi = -0.10074 \ \textrm{.}
\]

\noindent This means the reflectivity must be

\[
\eta = 0.990244 \ \textrm{.}
\]

\noindent Using these values however, the $\ket{0} + \ket{1}$ state is
prepared
only 1\% of the time.
We can get better efficiency by allowing the second
order term to be larger. For example choosing $\chi=-0.33714$ 
(corrosponding reflectivity $\eta = 0.91985$) the
coefficient of the
second order $\ket{2}$ term is 10 times smaller (probability 100 times
smaller) and now the state is prepared 8\% of the time.

{\it Superposition Basis Measurements}
Now that the superposition state $\ket{0}+\ket{1}$ has been produced
we also need to be able to measure in the basis state spanned by this
and the orthogonnal state $\ket{0} - \ket{1}$.  The device which is used to
perform measurements
in this
basis is a 50:50 beamsplitter with a known, positive superposition
state injected into one port and the unknown superposition injected
into the other, as shown in Fig.3.
%
\begin{figure}[ht]
   \begin{center}
   \includegraphics[width=0.65\linewidth]{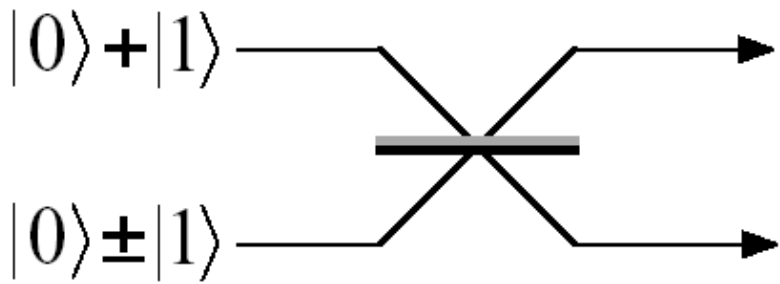}
   \end{center}
   \caption{Schematic for a device which performs measurements in the 
	superposition basis $\ket{0}\pm \ket{1}$.  The state to be measured 
	is in the lower mode while the upper mode contains the state 
	produced by the device described in figure~\ref{bs}.  
	$\ket{0} + \ket{1}$ is detected with certainty when one and only one 
	photon is measured in the top mode and $\ket{0} - \ket{1}$ when one photon 
	is found only in the lower mode.}
   \label{SB}
\end{figure}

\noindent One finds the state for a positive phase superposition to be

\[
\frac{1}{2}\ket{00} + \sqrt{\frac{1}{2}}\ket{10} +
\sqrt{\frac{1}{8}}\left(\ket{20} - \ket{02}\right)
\]

\noindent and for a negative phase

\[
\frac{1}{2}\ket{00} + \sqrt{\frac{1}{2}}\ket{01} -
\sqrt{\frac{1}{8}}\left(\ket{20} - \ket{02}\right)
\]

\noindent If one photon is measured in the
first mode then the unknown superposition had positive phase whilst a
one in the second
mode indicates a negative phase superposition. No descision can be
made if zero photons are counted in both modes or if two photons are
found in one. The measurement
succeeds one half of the time.

{\it Hadamard Gate} Now we have enough tools to perform the task that is
desired.  Fig.4 shows the construction of a single rail Hadamard gate.
%
\begin{figure}[ht]
   \begin{center}
   \includegraphics[width=0.9\linewidth]{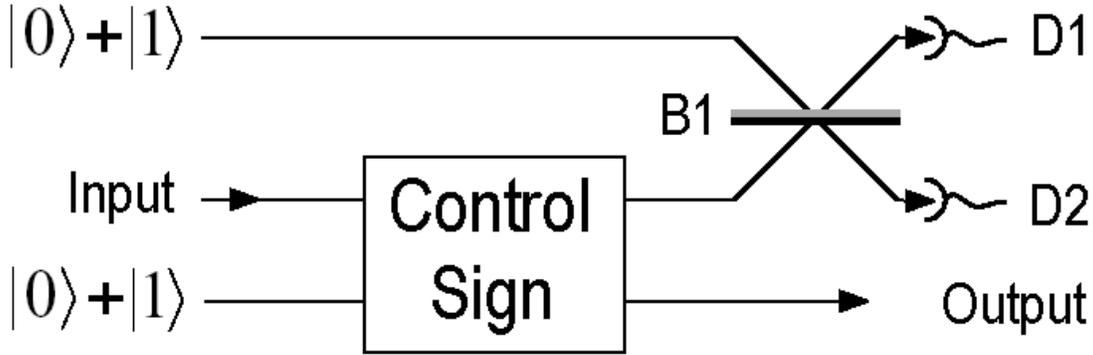}
   \end{center}
   \caption{Schematic of Hadamard gate constructed using devices from 
	figures~\ref{CS}, \ref{bs} and~\ref{SB}.  The $\ket{0} + \ket{1}$ 
	states are assumed to be created by a device similar to that in 
	figure~\ref{bs}.}
   \label{HG}
\end{figure}

\noindent This gate requires two $\ket{0}+\ket{1}$ superpositions, one
control sign gate (as described in the previous section) and one
$\ket{0}\pm\ket{1}$ measurement.  If the input is arbitrary
(i.e. $\alpha\ket{0}+\beta\ket{1}$) then the bottom two modes at the far
left can be
described by the state

\[
\ket{\Phi} = \alpha\ket{00} + \alpha\ket{01} + \beta\ket{10} + \beta\ket{11}
\]

\noindent where the first number is the number state of the upper mode.
After the control sign operation the state looks like

\[
\ket{\Phi} = \alpha\ket{00} + \alpha\ket{01} + \beta\ket{10} - \beta\ket{11}
\]

\noindent Now a $\ket{0}\pm\ket{1}$ measurement is made on the top mode
from the control sign gate.  If we only take the positive result then the
output state in the output mode is

\[
\ket{\psi} = \alpha(\ket{0} + \ket{1}) + \beta(\ket{0} - \ket{1})
\]

\noindent which is Hadamard logic. Noting that differential propagation
produces a phase rotation

\begin{eqnarray*}
\ket{\psi;t} &=& e^{-i\omega \hat{a}^\dagger\hat{a} t}
(\ket{0}+\ket{1}) \\
&=& \ket{0} + e^{-i\omega t}\ket{1}
\end{eqnarray*}

\noindent Allows us to then apply arbitrary rotations and thus
completes our universal set of gates.

\section{Testing the Gates}

\begin{figure}[ht]
   \begin{center}
   \includegraphics[width=0.7\linewidth]{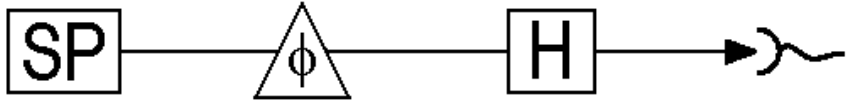}
   \end{center}
   \caption{Schematic of experimental test modeled to simulate the 
	functionality of these gates using realistic detectors. `SP' denotes 
	the $\ket{0}+\ket{1}$ production device (Fig~\ref{bs}), $\phi$ 
	performs a phase shift of the type $\ket{0}+e^{i\phi}\ket{1}$ and 
	`H' deontes a Hadamard gate as shown in figure~\ref{HG}.} 
   \label{EP}
\end{figure}

In our discussion so far we have assumed unit detector efficiency and
the ability to differentiate between zero, one and two photons. Off
the shelf photon counters on the other hand have efficiencies of
around 65\% and can only differentiate between zero and more photons.
This would be insufficient for demonstrating our single rail gates.
However state of the art detectors have recently been
described \cite{tak99} with 90\% efficiency and the ability to
differentiate between zero, one or two photons. We now model the
performance of a
simple single rail test circuit if implemented with these state of the
art detectors. The test circuit is shown in Fig~\ref{EP}. 
%

\noindent In the figure `SP' produces the state
$\ket{0}+\ket{1}$ non-deterministically, `H' is a Hadamard gate,
$\phi$ is a phase shift
and there is a detector after the Hadamard gate.  After the phase shift,
the state looks like (unnormalised)

\[
\ket{\psi} = \ket{0} + e^{i\phi}\ket{1} \ \textrm{.}
\]

\noindent This state is passed through the Hadamard gate, so the output
state at the detector is (unnormalised)

\[
\ket{\psi} = (1 + e^{i\phi})\ket{0} + (1-e^{i\phi})\ket{1} \ \textrm{.}
\]

\noindent So as the phase shift $\phi$ is changed the superposition at the
output changes from $\ket{0}$ to $\ket{1}$ over the range of $\phi = 0$ to
$\phi = \pi$, ideally with 100\% visibility.

\begin{figure}[ht]
   \begin{center}
   \includegraphics[width=\linewidth]{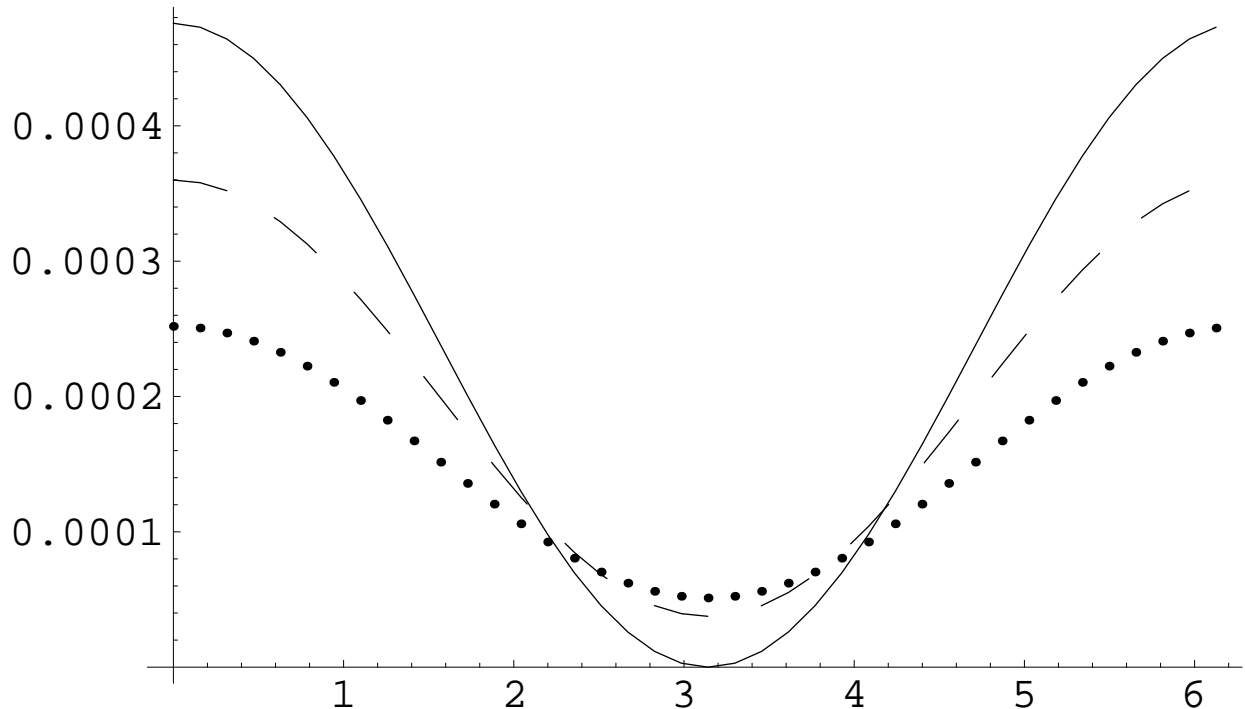}
   \end{center}
   \caption{Plot of detection probability verses the variation of the phase
	$\phi$ in figure~\ref{EP}.  The solid line is a simulation with 
	100\% efficient detectors, dashed line with 90\% efficient detectors
	and the dotted line with 80\% efficient detectors.  The visibilities 
	of these fringes are 1.00, 0.91 and 0.66 respectively.}
   \label{PG}
\end{figure}

Photon loss in inefficient detectors is included into the analysis of the
gate by modelling the inefficient detector as a beamsplitter of
transmittivity equivalent to the efficiency of the detector followed by
a perfect detector.  The reflected mode is traced over.
Each of the detectors used in the construction of the Hadamard gate and the
$\ket{0}+\ket{1}$ state production devices was simulated by this technique.
Figure~\ref{PG} shows the results of this simulation. The $\ket{0}+\ket{1}$ 
state producer (Fig~\ref{bs}) use a coherent strength of 
$\chi=-0.33714$ and by equation~\ref{rf} a reflectivity of 
$\eta=0.91985$.
Shown is the
probability of detecting no photon in the output, given the correct
combination of conditioning detector results.
The probability is not normalised
to this conditioning, so the y-axis reflects the probability of obtaining
this event (including the conditioning probability) for a single run.
The x-axis is the size of the phase shift. The phase shift
is relative to the the phase of the the coherent states used to
produce the $\ket{0}+\ket{1}$ superposition states used as an input and
as gate resources.

%
\noindent This figure shows detectors of 100\% efficiency (solid line),
90\% efficiency (dashed line) and 80\% efficiency (dotted line) using the
photon loss model described above.  The visibility of these curves is
unity for 100\%, 0.81 for 90\% and 0.66 for 80\% efficient detectors.

\section{Conclusion}

We have shown that a universal set of gates can be implemented 
non-deterministically for the qubit encoding in which the vaccum state, 
$\ket{0}$, represents logical zero 
and the single photon state, $\ket{1}$, represents logicial one. The 
implementation uses only linear optics and photo-detection. It was also 
shown that a demonstration of the operation of these gates is plausible using  
current detector technology.

\end{document}